\documentclass[a4paper,10pt]{article}
\usepackage[utf8]{inputenc}
\usepackage{braket}

 \title{Holographic Complexity and 
Fidelity Susceptibility as Holographic Information Dual to Different  Volumes in AdS }
\author{ N. S. Mazhari$^{1}$, Davood Momeni$^{1}$ ,\\ Sebastian Bahamonde$^{2}$, Mir Faizal$^{3,4}$, Ratbay Myrzakulov$^{1}$ 
\\\\$^1$ Eurasian International Center for Theoretical Physics \\
and Department of General Theoretical Physics, \\
Eurasian National University, Astana 010008, Kazakhstan\\$^2$
Department of Mathematics, University College London,\\  Gower Street, London, WC1E 6BT, UK 
 \\$^3$ Irving K. Barber School of Arts and Sciences, 
 \\ University of British
 Columbia - Okanagan,  
 3333 University Way,\\  Kelowna,   British Columbia, V1V 1V7, Canada
\\$^4$ Department of Physics and Astronomy, University of Lethbridge, \\
Lethbridge, Alberta, T1K 3M4, Canada
}
\date{}
\begin{document}

\maketitle
\begin{abstract}
 The holographic complexity and 
fidelity susceptibility have been defined as new quantities dual to different volumes in AdS. 
In this paper, we will use these new proposals to calculate both of these quantities for a variety of interesting 
 deformations of AdS. 
 We obtain the  holographic complexity and 
fidelity susceptibility for an AdS black hole, Janus solution and a solution with cylindrically symmetry, an  
inhomogeneous background and  a hyperscaling violating background. It is observed that the holographic complexity 
depends on the size of the subsystem for all these solutions and the fidelity susceptibility does not have any such dependence.  
\end{abstract}

\section{Introduction}
The information theory  deals with the ability of an observer to process relevant information, and it is important as studies done in  
 different branches  of  physics seem to  indicated that  the laws of physics are informational theoretical processes \cite{info, info2}.  
 It is important to know how much information is lost, when an observer processes the relevant information, 
 and it is also important to quantify this abstract concept relating to the of loss of information in a process. 
 The quantity which quantifies this concept relating to the loss of information is the entropy, 
 and it is one of the   most important quantities in information theory. 
 As the laws of physics can be represented by informational theoretical processes, entropy has been used 
 to analyse the behavior of   physical systems ranging from   condensed matter physics to gravitational physics.
 It may be noted that in Jacobson formalism, it is even possible to obtain  the  general relativity from  thermodynamics 
 \cite{z12j, jz12}. Thus, it is possible that the geometry of spacetime is 
  an emergent structure, and it  emerges  due to 
 an information theoretical process. In the Jacobson formalism it is important to assume a certain scaling 
 behavior of entropy to obtain general relativity, i.e., the  
 maximum entropy of a region of space scales with its  
area, and this has been motivated from the physics of black holes. This is because the black holes are maximum entropy objects, 
and the entropy of a black hole scales with its area. The holographic principle is motivated from this 
observation that the maximum entropy of a region of space scales with its 
area  \cite{1, 2}. The holographic principle states that the number of degrees of freedom in a region of space is equal to the number of degrees 
of freedom on the boundary surrounding that region of space. The AdS/CFT correspondence is  of the most
important realizations of the holographic principle \cite{M:1997}, and  it relates the supergravity solutions in  AdS spacetime
to the superconformal field theory on the boundary of that AdS spacetime. 

It is interesting to note that the holographic principle which was initially proposed due to the scaling behavior of entropy in 
black holes, may also lead to a solution of the black hole information paradox. The black hole information paradox occurs due to the observation
that  classically information cannot get of a black holes  and black holes  evaporate due to Hawking radiation. 
This is because it has been proposed that   quantum entanglement can be used to analyse the     microscopic  
of a black hole, and it is hoped that this  may resolve the black hole information paradox
\cite{4,5}.
The AdS/CFT correspondence, which is a concrete regularization of the holographic principle,
can be used to  quantify   quantum 
entanglement in terms of the holographic
entanglement entropy .  The holographic entanglement entropy 
 of a CFT   is dual to the area of a minimal surface defined in the bulk of an 
 asymptotically AdS spacetime. Now  for a  
   subsystem $A$ (with its complement),  $\gamma_{A}$ can be defined as  
   the $(d-1)$-minimal surface extended 
into the AdS bulk with the boundary $\partial A$. So, the holographic 
entanglement entropy for this subsystem, can be written as \cite{6, 6a}
\begin{equation} \label{HEE}
S_{A}=\frac{Area(\gamma _{A})}{4G_{d+1}}
\end{equation}
where $G$  is the gravitational constant in the AdS spacetime. 

It is important to know how much information is retained in a system, but it is also  important to know, how easy is it 
for an observer to process this information. Just as entropy quantifies  the abstract idea of loss of information, complexity 
quantifies the abstract idea of the difficulty to process this information, and so just like entropy,  complexity 
is a fundamental quantity relating to 
information theoretical processes. As the laws of physics can be represented in terms of informational theoretical processes, it 
is expected that complexity can be viewed as another fundamental physical quantity, and it is expected that laws 
of physics should be written in terms of complexity. It is interesting to note that
 complexity (like entropy) has been used to study condensed matter systems \cite{c1, c2} and  molecular physics \cite{comp1}. In fact, 
 complexity is also important in quantum computing \cite{comp2}. Complexity is also important in analysis the physics of black holes, 
 as it has been recently  proposed that the information may not be ideally lost in a black hole, but it may be 
lost for all practice purposes as it would be impossible to reconstruct it from the Hawking radiation \cite{hawk}. 
However, unlike entropy, there is no universal definition of complexity  of a   system, and there are different 
proposals to define the complexity of any systems.   However,  it is possible to define 
complexity   holographically. In fact, recently 
 holographic complexity has been  defined as a quantity dual to   a volume of  codimension one time 
slice in anti-de Sitter (AdS) \cite{Susskind:2014rva1,Susskind:2014rva2,Stanford:2014jda,Momeni:2016ekm}, 
\begin{equation}\label{HC}
Complexity = \frac{V }{8\pi R G_{d+1}},
\end{equation}
where $R$ and $V$ are the radius of the curvature and the volume in the AdS bulk. 

The different proposals for complexity could be related to the 
different possible ways to 
define this volume in the bulk. It is possible to define   complexity as dual to the 
  maximal volume in the AdS which
ends on the time slice at the AdS boundary $V = V_{max}$ \cite{r5}, and it has been   demonstrated that this proposal corresponds to the 
fidelity susceptibility   of the boundary CFT. Hence, this quantity is called   fidelity susceptibility  even in the bulk, and we will denote it by 
$\Delta\chi_F$. 
It is interesting to note that the fidelity susceptibility of the boundary theory can be used for 
  analyzing   the quantum phase transitions \cite{r6,r7, r8}, and thus it is possible to study quantum phase transitions holographically. 
However, it is also possible use a subsystem $A$ with its complement, and define the volume as  $V = V(\gamma)$. This volume is the 
  volume enclosed by the   minimal surface used to calculate the holographic 
entanglement entropy \cite{Alishahiha:2015rta}, and it can also be used to holographically define complexity and will be denoted by 
$\Delta \mathcal{C}$.  As we want to differentiate 
it from the case, where the maximum volume has been used to calculate the complexity of a system, 
we shall call it holographic complexity (this terminology 
follows from \cite{Alishahiha:2015rta}, where such a quantity is called holographic complexity).
Thus, 
in this paper, the maximum volume of a system $V = V_{max}$ will be used to calculate the
fidelity susceptibility, and the $V = V(\gamma)$ will be used
to calculate the holographic complexity of such a system. As  complexity is a new physical 
quantity and it is expected that laws of physics can be written in terms of complexity, 
we will use these recent proposals to calculate the holographic 
complexity and  fidelity susceptibility for various deformed  AdS solution. 

As it has been proposed that the 
holographic complexity and fidelity susceptibility of a boundary theory  can be holographically 
calculated from a deformed AdS bulk solution, it would be interesting to calculate such 
quantities for AdS bulk solutions which have interesting boundary dual solutions.  
These quantities calculated in the bulk 
could  be used to understand the behavior of  the boundary field theory dual 
to such geometries. This is the main motivation to study such quantities for an 
AdS$_{d+2}$ black hole, Janus solution, cylindrical solution,   inhomogeneous backgrounds, and  
  hyperscaling violating backgrounds. Most of these deformed AdS solutions have interesting 
 boundary dual.  In this paper, we will also mention some interesting 
 field theories which are dual to these 
 deformations of the AdS spacetime. Thus, it is important to analyze such quantities in the bulk 
to possible  understand their behavior  in the boundary field theory dual to such a bulk.  
 Furthermore, apart from having interesting boundary duals, these solutions are interesting 
 geometric solutions. So, by calculating these quantities for these solutions, we will also
 try to understand certain universal features of holographic complexity and fidelity susceptibility
 for different deformations of the AdS geometry.

  The organization of this paper is as follows: In Sec. \ref{excitedd}, we will 
 study the holographic quantities
  for an AdS$_{d+2}$ black hole. In Sec.~\ref{Januss} and Sec.~\ref{cylin} we will
  examine the Janus 
  and cylindrical solutions and we will show that the holographic complexity is 
  different than the fidelity susceptibility, 
  which is an opposite result as it was given in previous works. In Sec.~\ref{inhom}
 and Sec.~\ref{hypers}, as two complementary and interesting examples, 
 the holographic complexity and the  fidelity susceptibility will be also
 studied for geometries with inhomogeneous and 
 hyperscaling violating backgrounds respectively. 
 Finally, we will conclude our main results in Sec. \ref{conclusion}.

\section{AdS Black Holes}\label{excitedd}
In the holographic picture, an excited state in CFT on the boundary is dual to  a deformation of AdS in the bulk.   
This deformed metric could be expressed  asymptotically by an AdS geometry. 
Such AdS black hole can be used to holographically model superconductor \cite{super, super2}. 
It is important to understand the behavior of fidelity susceptibility for superconductors. 
In fact, the fidelity susceptibility in topological superconductors  has been obtained, and this was done 
by solving Bogoliubov-de Gennes equations \cite{fedl}. As it is possible to holographically describe superconductors using 
AdS black holes, it will be possible to obtain the fidelity susceptibility for such  field theories which are boundary dual to AdS black holes, 
by calculating the fidelity susceptibility for   AdS black holes. So, we will calculate the fidelity susceptibility for a AdS black hole, 
and this will be dual to the maximum volume. However, we will also use a subsystem, and calculate the holographic complexity for such 
AdS black holes. 

We will use a  deformed Poincare metric for AdS$_{d+2}$  black hole to perform this calculation, and this metric can be written as  
\begin{eqnarray}
ds^2=\frac{R^2}{r^2}\left(-h(r)  dt^2+\frac{dr^2}{h(r)}+d\rho^2+\rho^2 d\Omega_{d-1}^2\right)\,.  \label{metric}
\end{eqnarray}
By setting the metric function $h(r)=1$, we can recover a pure AdS space-time. This metric function   now gets deformed as 
 $h(r)=1-m r^{d+1}$ where $m$ is a constant. In analogy with the  pure AdS background, 
 subsystem in the bulk can 
 be parametrized by $\rho= f(r)$. However,  in this case, the function $f(r)$ does not have a closed simple. 
 The minimal hypersurface can be obtained by minimizing  the auxiliary functional, 
\begin{eqnarray}
&&Area=\Omega_{d-1}\int dr\Big(\frac{R}{r}\Big)^d f(r)^{d-1}\sqrt{f'(r)^2+\frac{1}{h(r)}}\,.\label{action}
\end{eqnarray}
Here, prime denotes derivative with respect to the radial coordinate $r$ and $\Omega_{d-1}=2\pi^{d/2}/\Gamma(d/2)$. 
The appropriate boundary conditions for this system are $f(0)=r_t $ and $ f'(0)=0$, where $r_t$ denotes the classical turning point of $f(r)$. 
The associated  equation of motion obtained from this action  can be expressed as 
\begin{eqnarray}
&&f''+\frac{1-d}{f}\Big(f'(r)^2+\frac{1}{h(r)}\Big)+\frac{f'}{2}\frac{h'}{h}=0\label{eom}\,.
\end{eqnarray}
If $m$ is sufficiently small, we can write the solution to this equation up to first order in $m$ as an expansion of $f(r)$ as follows 
\begin{equation}
 f(r)=f_0(r)+mf_1(r)+{\cal
O}(m^2)\,, 
\end{equation}
where $f_0(r)$ is the exact solution of Eq. (\ref{eom}). For $h(r)\approx 1$,  the initial
conditions are given by 
\begin{eqnarray}
 f_0(r)=r_t\,, \ \ \ \ f_0'(r=0)=0\,,
\end{eqnarray}
and hence we obtain 
 \begin{equation}
 f_0(r)=\sqrt{r_t^2-r^2}\,.
 \end{equation}
The profile of the minimal surface at leading order in $m$ can be written as
\begin{eqnarray}
  m\ell^{d+1}\ll 1\,.
 \end{eqnarray}
Thus, we can write the metric function as
 \cite{Bhattacharya:2012mi} 
\begin{eqnarray}
f(r)=\sqrt{r_t^2-r^2}\bigg(1+\frac{2 r_t^{d+3}-r^{d+1}(r_t^2+r^2)}{2(d+2)(r_t^2-r^2)}m\bigg)+{\cal
O}(m^2)\,,\label{f(r)}  
\end{eqnarray}
where we assumed a regularity at $r = r_t$. The parameter $r_t$ is a free positive constant which is related to the radius $\ell$  
of the subsystem  by 
\begin{eqnarray}
&&\ell=2\int_{0}^{r_t}dr\Big(\frac{r}{r_t}\Big)^d\displaystyle\sqrt{\frac{1}{h(r)\big(1-\big(\frac{r}{r_t}\big)^{2d}\big)}}\,.
\end{eqnarray}
The length of the entangled system is fixed, so that we can compute the 
 turning point $r_t$ to leading order in $m$, yielding 
\begin{eqnarray}\label{tp}
r_t=\ell\left(1-\frac{m\ell^{d+1}}{d+1}+{\cal O}\bigg((m\ell^{d+1})^2\bigg)\right)\,.
\end{eqnarray}
The volume of codimension one spacetime enclosed by the minimal area   is defined by the following integral, 
\begin{eqnarray}
  V (\gamma) =\frac{\Omega_{d-1} R^{d+1}}{d}\int_\varepsilon^{r_t} dr\,  \frac{f(r)^d}{r^{d+1}\sqrt{h(r)}}\label{volume}\,.
\end{eqnarray}
Here $\epsilon$ denotes a UV cut-off. By substituting Eq.~(\ref{f(r)}) into the above equation and then by evaluating the integral, we obtain 
\begin{equation}\label{Delta V}
\Delta V=V_{BH}-V_{AdS_{d+1}}=\frac {4 a_{{d}}{R}^{d+1}\Omega_{d-1}}{ \left( d+2 \right) d} \left(m \ell^{d+1}\right)\,, 
\end{equation}
where the coefficients $a_d$ are defined as  
 \begin{eqnarray}\label{rho'}
&&a_d=
\left\{
\begin{array}{lr}\  A & d=2n,n\in\mathcal{Z},  \\

\  B & d=2n+1,n\in\mathcal{Z}\,, %
\end{array}%
\right. 
\end{eqnarray}%
where 
\begin{eqnarray}
 A = 
\sum_{p=0}^{\frac{d}{2}} \sum _{q=0}^{\frac{d}{2}-1} {\frac{d}{2}\choose p}{\frac{d}{2}
-1\choose q} \Big[( d+2 )( q+\frac{3}{2})( 
q-d/2)( q+\frac{1}{2})( -1) ^{p} \\\nonumber + ( p+
\frac{1}{2})(( q+1) d+2q+\frac{3}{2}) d( -1
 ) ^{q}\Big] \\ \times \Big[ ( 2p+1)( 2q-d ) 
( 2q+3)( 2q+1) \Big]^{-1}, \nonumber \\ 
B = \sum _{p=0}^{\infty} \sum _{q=0}^{\infty} {\frac{d}{2}\choose p}{\frac{d}{2}
			-1\choose q} \Big[( d+2 )( q+\frac{3}{2})( 
q-d/2)( q+\frac{1}{2})( -1) ^{p}\\ \nonumber  + ( p+
\frac{1}{2})(( q+1) d+2q+\frac{3}{2}) d( -1
 ) ^{q}\Big]  \\ \nonumber \times \Big[ ( 2p+1)( 2q-d ) 
( 2q+3)( 2q+1) \Big]^{-1}, 
\end{eqnarray}
where ${\frac{d}{2}\choose p}$ denotes the binomial coefficient. Now, by using Eqs.~(\ref{Delta V}) and (\ref{HC}) we 
can obtain the holographic complexity, which is 
\begin{eqnarray}
&&\Delta\mathcal{C}=\frac {4 a_{{d}}{R}^{d}\Omega_{d-1}}{8\pi G \left( d+2 \right) d} \left(m \ell^{d+1}\right)\,.
\end{eqnarray}
It may be noted that this expression for $\Delta\mathcal{C}$ is different from the holographic complexity calculated  
  in \cite{Alishahiha:2015rta}, which was given by 
\begin{eqnarray}
\Delta\mathcal{C}=\frac{c_dR^d \Omega_{d-1}}{8\pi Gd}(m\ell^{d+1})^2\,.\label{ali}
\end{eqnarray}
Therefore, $\Delta C \propto m \ell^{d+1}$, and not $(m \ell^{d+1})^2$ as was proposed in  \cite{Alishahiha:2015rta}.

Now, we will calculate the fidelity susceptibility for an deformed AdS state with metric (\ref{metric}). To do this,  
we need to evaluate the $\mbox{Vol}(\Sigma_{max})$ 
for the metric given by Eq.~(\ref{metric}). Therefore, we can set $t=0$ and consider the codimension one hypersurface. 
In order to compute the volume integral, we can use the expression (\ref{volume}) with different integral limits, i.e.,
changing $r_t$ by the horizon $r_{+}$ 
given by $h(r_{+})=0$, and $r_{+}=m^{-\frac{1}{d+1}}$. Thus,    
the volume term   can be  expressed as \cite{r5}, 
\begin{eqnarray}
\Delta \mbox{Vol}(V_{max})=\frac{b_dR^{d+1}V_d }{d}(m^{d/(d+1)})\,,
\end{eqnarray}
where $V_d={\mbox{Vol}}\{V_d:d\rho^2+\rho^2d\Omega_{d-1}\}$. The fidelity susceptibility can be written as 
\begin{eqnarray}
&&\Delta\chi_F(\lambda)=n_d\frac{b_dV_d }{d}(m^{d/(d+1)})\,.\label{fidelity}
\end{eqnarray}
This fidelity susceptibility can be used to obtain the fidelity susceptibility for the holographic superconductors. 
However, we have also demonstrated that it is also possible to derive other quantities dual to 
the volume in the bulk, and this is the holographic complexity. 
It may be noted that   the 
holographic complexity depends on the size of the subsystem  $\ell$ and
as the fidelity susceptibility was calculated for the full system, no such dependence has been observed. 
Furthermore, as holographic entanglement entropy has been calculated for AdS black holes \cite{ee12, ee22}, 
and the holographic complexity is calculated using the same surface as entanglement entropy, so  we have calculate both 
holographic complexity and fidelity susceptibility for AdS black holes.  
\section{ Janus Solution}\label{Januss}
It is possible to obtain a 
nonsupersymmetric dilatonic deformation of AdS geometry 
as an exact nonsingular solution of the type IIB supergravity \cite{Bak:2003jk}. 
The   gauge theory dual to this solution  has a different Yang-Mills coupling in each of the two halves of the boundary 
spacetime divided by a codimension one defect. The  
  structure of the boundary  
  and   the string configurations corresponding to Wilson loops  for this solution have been studied \cite{Bak:2003jk}. 
  This solution is called the Janus solution, 
and it has also been possible to study the  supersymmetric Janus solution \cite{Freedman:2003ax}. 
It has been demonstrated that the Janus solution has   quantum level conformal symmetry, and this was done by 
using  conformal perturbation theory to study various correlation functions \cite{j1}. 
The holographic entanglement entropy in the presence of a conformal interface   has been 
recently calculated,  and  it  was observed that for   the supersymmetric Janus solution the holographic entanglement entropy calculated from the bulk  was in   exact agreement  with the  calculations done using a CFT \cite{entr}. 
As the holographic complexity is calculated using the same surface as the holographic entanglement entropy, 
we will calculate the holographic complexity for the Janus solution. We will also calculate the 
fidelity susceptibility for the Janus solution, and this can be used to understand the behavior of 
fidelity susceptibility for a system of consisting of 
different Yang-Mills coupling in each of the two halves of the boundary.   

It is possible to use the AdS$_2$ slice of the deformed AdS$_3$ has been used to obtain   the Janus solution. This solution  
 is an exact solution defined using  the following Euclidean bulk action,
\begin{eqnarray}
&&S=-\frac{1}{16\pi G_N}\int d^3x\sqrt{g_{E}}\Big(R-\phi_{;\mu}\phi^{;\mu}+\frac{2}{R^2}\Big)\,.
\end{eqnarray}
Here, $\phi$ is the massless bulk scalar field. The metric of the Janus solution and the profile of dilaton field, is given by the Euclidean metric
\begin{eqnarray}
&&ds^2=R^2(dy^2+\frac{f(y)}{z^2}(dz^2+dx^2))\label{Janus}\,,\ \ \phi(y)=\gamma\int_{-\infty}^{y}\frac{dy}{f(y)}+\phi_1\,,\\&&
f(y)=\frac{1}{2}\big(1+\sqrt{1-2\gamma^2}\cosh(2y)\big)\,,\ \ \gamma\leq \frac{1}{2}\,,\,\, \phi_1=\phi(-\infty)\,.
\end{eqnarray}
For this geometry, the coupling constant for the ground state $|{\Omega_{1}}>$ is dual to $\phi_{1}$.
The fidelity susceptibility was computed in  \cite{r5}, and is given by 
\begin{eqnarray}
&\Delta\chi_F(\lambda)=\displaystyle\frac{cV_1}{12\pi\epsilon}\,,\label{FJanus}
\end{eqnarray}
where $V_1$ is the volume of the AdS$_2$ per unit radius and $\epsilon$ is a UV cutoff.
Now, we will compute the holographic complexity for the Janus solution represented by the metric (\ref{Janus}).
The area functional for an entangled region $A=\{x\in[0,L],z=z(y)\}$ will be given by
\begin{eqnarray}
&&Area=R^2L\int_{-y_{\infty}}^{y_{\infty}}dy\sqrt{\frac{f(y)}{z^2}\Big(1+\frac{f(y)}{z^2}z'^2\Big)}\,,\ \ \ z'\equiv \frac{dz}{dy}\,.\label{AJanus}
\end{eqnarray}
Moreover, the entangled length and volume are
\begin{eqnarray}
 \ell&=&2\int_{0}^{y_t}z'(y)dy\,,\\
V(\gamma)&=&R^3L\int_{-y_{\infty}}^{y_{\infty}}dy f(y)\int_{\epsilon}^{z(y)}\frac{dz}{z^2}\,,
\end{eqnarray}
which could be simplified by subtraction of the pure AdS portion from the AdS black hole. Thus, we can write the 
finite part as 
\begin{eqnarray}
&&V(\gamma)=-R^3L\int_{-y_{\infty}}^{y_{\infty}}dy\frac{ f(y)}{z(y)}\,.
\end{eqnarray}
Next, we need to find $z(y)$ which minimizes the area functional (\ref{AJanus}) subject to the boundary conditions $z(0)=z_t$ and $z'(0)=0$. 
We can expand $z(y)=z_{0}+z_{1}y^2+z_{2}y^4$ in series to find its solution up to fourth order in $y$, 
\begin{eqnarray}
z \left( y \right) &=& z_{{t}}-\frac{z_{{t}}}{2}\,{\frac { \left( 1+\sqrt {1-2\,
{\gamma}^{2}} \right) {y}^{2}}{\sqrt {1-2\,{\gamma}^{2}}+1-{\gamma}^{2
}}} \\ \nonumber && -\frac{z_{{t}}}{12}\,{\frac { \left( -9-9\,\sqrt {1-2\,{\gamma}^{2}}+17\,{
\gamma}^{2} +8\,{\gamma}^{2}\sqrt {1-2\,{\gamma}^{2}} \right) {y}^{4}}{
 \left( \sqrt {1-2\,{\gamma}^{2}}+1-{\gamma}^{2} \right) ^{2}}}\\ \nonumber && +\mathcal{O}(y^6)\,.
\end{eqnarray}
Using this solution which is valid near the Cauchy surface $y=0$, we can evaluate the length $\ell$ and from it. Now we 
obtain numerically that $z_t\approx \ell^{1/3}+O \left( {\gamma}^{2}
 \right)$. Finally,
 the  holographic complexity is given by the following expression:
\begin{eqnarray}
\Delta\mathcal{C}_A &=& \frac{1}{8\pi R G_{d+1}}\Big( 9.114502677\,{z_{{t}}}^{-1}+{\frac {16}{9}}\,{\frac {y_{{\infty }}}{z
_{{t}}}}+\mathcal{O}\left( {y_{{\infty }}}^{-1} \right)\nonumber \\ &&  +\mathcal{O} ( {\gamma}^{2})\Big)\,.\label{CJanus}
\end{eqnarray}
It is remarkable to see that (\ref{FJanus}) and (\ref{CJanus}) are   different even in the first orders. 
It can be noted that in the leading order of expansion, 
\begin{eqnarray}
\Delta\mathcal{C}_A\approx \ell^{-1/3}\,. 
\end{eqnarray}
Thus, we have obtained an expression for the holographic complexity and fidelity susceptibility for Janus solution. 
It may be noted that as the holographic complexity is calculated for a subsystem, it depends on the size of the subsystem. 
However, the expression (\ref{FJanus})   is independent of the entangled length, as the fidelity susceptibility  is calculated 
for the full system. 
Furthermore, the fidelity susceptibility for this solution can be used to understand the behavior of fidelity susceptibility 
for a system described by two different Yang-Mills coupling in each of the two halves of the boundary.

\section{Cylindrical Symmetry}\label{cylin}
In order to examine the properties of the holographic complexity and the fidelity susceptibility, 
it is important to study geometries with different types of symmetries as for example, cylindrical ones.
A very interesting cylindrically symmetric solution was presented in \cite{Miradhayeejafari:2009tu},
where a massless scalar field minimally coupled to gravity with cosmological constant was obtained. 
This solution can be understood as a generalization of the Buchdahl's solution without cosmological
constant and the Levi-Civita-$\Lambda$ solution without a scalar field. Cosmologically speaking, 
it was also showed in \cite{Miradhayeejafari:2009tu} that this solution can describe a Cyclic universe  
  in a braneworld model.
   The   Einstein-Rosen waves and the self-similarity hypothesis has been 
   studied using cylindrical symmetric solution \cite{cyli1}. It has been observed that such    
   solutions are reduced to part of the Minkowski spacetime with a  conically singular axis 
   if the homothetic vector is orthogonal to the cylinders of symmetry. 
  A vortex line solution for Abelian Higgs field has also been analysed using 
  a cylindrical symmetric solution \cite{cyli}.    
  In this study, it  was demonstrated  that  the mass density of the string is uniform and 
    dual to the discontinuity of a logarithmic derivative of correlation function of the boundary scalar operator. 
It would be interesting to analyse the fidelity susceptibility for a cylindrical symmetric solution, as this 
can be used to understand the behavior of fidelity susceptibility for the Abelian Higgs field. So, now we will calculate 
the fidelity susceptibility for a cylindrical symmetric solution. We will also calculate the holographic complexity 
for such a solution. 
 
To  calculate the  holographic complexity and the 
fidelity susceptibility for a solution with cylindrical symmetry, we will use a cylindrical analogous of the AdS$_4$ \cite{Miradhayeejafari:2009tu}.
This solution is obtained from the action of a massless scalar field $\phi$ 
in the presence of a cosmological constant, i.e., the action 
\begin{eqnarray}
&&S=-\frac{1}{16 \pi G_N}\int \sqrt{-g}d^4x \Big(R-2\Lambda+\phi_{;\mu}\phi^{;\mu}\Big)\,.
\end{eqnarray}
In Weyl cylindrical coordinates $x^{\mu}=(t,r,\varphi,z)$, 
the field equation given by $R_{\mu\nu}+\Lambda g_{\mu\nu}=\phi_{;\mu}\phi_{;\nu}$ has the following exact solution for the metric and the scalar field:
\begin{eqnarray}
ds^2&=&dr^2+e^{-2\sqrt{\frac{-\Lambda}{3}}r}(\xi^2e^{-2\sqrt{-3\Lambda}r}+1)^{2/3}(-dt^2+d\varphi^2+dz^2)\label{cylind}\,,\\
\phi&=&\pm\frac{2\sqrt{6}}{3}\tan^{-1}(\xi e^{-\sqrt{-3\Lambda}r})\,.
\end{eqnarray}
Here, $\xi$ is a scalar field parameter which determines the curvature strength of the scalar field. If this parameter is complex, this solution has 
a naked singularity whereas if  $|\xi|>1$ it does not have any such singularity.
This solution reduces to  the cylindrical Levi-Civita-Lambda solution when $\phi=0$, 
and it reduces to the Buchdahl solution (the solution of Einstein gravity with massless scalar field)
when we set $\Lambda=0$. Following the proposal of \cite{r5},
to find the fidelity susceptibility, we need to evaluate the following integral 
\begin{eqnarray}
&&S(\xi)=\frac{R}{4\pi G_N}(2\pi L)\int_{-r_{\infty}}^{r_{\infty}}dr (1+\xi^2 e^{6r/\ell})^{2/3}\,,\label{sxi}
\end{eqnarray}
where we defined the AdS radius as $\ell^2\Lambda=-3$. The action (\ref{sxi}) evaluated at $\xi=0$ is
\begin{eqnarray}
&&S(0)=\frac{R}{4\pi G_N}(2\pi L)\int_{-\hat{r}_{\infty}}^{\hat{r}_{\infty}}d\hat{r}\,,\label{s0}
\end{eqnarray}
where $\hat{r}$ is obtained from the asymptotic form of the metric in the pure AdS case when $\xi=0$ given by ($r\to\infty$)
\begin{eqnarray}
&&ds_{pure}^2\sim d\hat{r}^2+e^{-2\frac{\hat{r}}{\ell}}(-dt^2+d\varphi^2+dz^2)\,.
\end{eqnarray}
Here $r_{\infty}$ is the one obtained from the asymptotic form ($r\to\infty$) of the metric in the massive AdS case when $\xi\neq0$, 
\begin{eqnarray}
&&ds_{massive}^2\sim dr^2+\xi^{4/3}e^{2r/\ell}(-dt^2+d\varphi^2+dz^2)
\end{eqnarray}
If we match two metrics, we find that 
\begin{eqnarray}
&&\xi^{2/3}e^{r_{\infty}{\ell}}=e^{\mp \hat{r}_{\infty}{\ell}}\,,\label{eq}
\end{eqnarray}
and then we need to consider two cases depending on the signs in the above equation.

Choosing the minus sign in Eq. (\ref{eq}, up to the second order of $\xi$, the difference of integrals (\ref{sxi}) and (\ref{s0}) gives us
\begin{eqnarray}
&&S(\xi)-S(0)=\frac{R L}{G_N}\Big(2r_{\infty}-\ell(1-\xi^{2/3})+\mathcal{O}(\xi^2)\Big)\,,
\end{eqnarray}
and using this expression, we can find the the fidelity susceptibility which it is given by 
\begin{eqnarray}
&&|<\Omega_{2}|\Omega_{1}>|\approx e^{S(\xi)-S(0)}\approx 1-\frac{LR}{G_N}\Big(1-\xi^{2/3}+\mathcal{O}(\xi^2)\Big)\,.
\end{eqnarray}

Choosing the plus sign in Eq. (\ref{eq}), following the same procedure as before, we find that the difference of the integrals are
\begin{eqnarray}
&&S(\xi)-S(0)=\frac{RL}{G_N}\Big(-\frac{2\ell}{3}\ln\xi+\frac{\ell\xi^2}{9}\sinh\big(\frac{6r_{\infty}}{\ell}\big)+\mathcal{O}(\xi^4)\Big)\,,
\end{eqnarray}
and then, the fidelity susceptibility becomes 
\begin{eqnarray}
&&|<\Omega_{2}|\Omega_{1}>|\approx e^{S(\xi)-S(0)}\approx \big(1+\frac{RL}{G_N}\frac{\ell\xi^2}{18}e^{\frac{6r_{\infty}}{\ell}}
\big)e^{-\frac{2R\ell L}{3G_N}\ln(\xi)\,
}\,.\label{fid}
\end{eqnarray}
 
 Now, to compute the holographic complexity for the metric (\ref{cylind}), we will suppose that the
entangled region is $\tilde{A}=\{r=r(\varphi),0<z<L,t=0,\varphi\in[0,\varphi_{\infty}]\}$, and so that the area functional is given by the following
\begin{eqnarray}
&&Area=2L\int_{0}^{\varphi_{\infty}} d\varphi\sqrt{f(f+r'^2)}\,,
\end{eqnarray}
where $f=e^{-2\frac{r}{\ell}}(\xi^2e^{\frac{6r}{\ell}}+1)^{2/3}$ and prime denotes differentiation
with respect to $\varphi$. Since the functional is not function of $\varphi$, the following first integral is a conserved quantity,
\begin{eqnarray}
&&
\frac{f^2}{\sqrt{f(f+r'^2)}}=E\,.
\end{eqnarray}
If we suppose that $r(0)=r_{t}$ and $r'(0)=0$, then $E=f(r_t)$, and hence we obtain 
\begin{eqnarray}
r'=\pm\sqrt{f\Big(\frac{f^2}{f(r_t)^2}-1\Big)}\,.
\end{eqnarray}
Therefore, the integral of area is minimized as follows:
\begin{eqnarray}
&&Area=2L\int _{0}^{r_{t}}dr\frac{f^2
}
{\sqrt{f(f^2-f(r_t)^2)}}\,,\label{A}
\end{eqnarray}
where $r_t$ is obtained from 
\begin{eqnarray}
&&\ell=2\int_{0}^{r_t}dr\frac{f}{\sqrt{f(f^2-f(r_t)^2)}}\,.\label{l}
\end{eqnarray}
In order to obtain the Holographic entanglement entropy, we need to solve (\ref{l}) to find $r_t$ and then replace it in (\ref{A}). 
For the holographic complexity, we need to evaluate the following integral
\begin{eqnarray}
&&V(\gamma)=2L\int_{0}^{\varphi_{\infty}} d\varphi\int_{r_t}^{r(\varphi)}fdr\,.
\end{eqnarray}
The minimal surface near the AdS horizon is given approximately by the following series expression:
\begin{eqnarray}
r \left( \varphi \right) &=&r_{{t}}- \left( {{\rm e}^{-2\,{\frac {r_{{t}}}{
\ell}}}}-{\xi}^{2}{{\rm e}^{4\,{\frac {r_{{t}}}{\ell}}}} \right) {\varphi}^{2}{
\frac {1}{\sqrt [3]{1+{\xi}^{2}{{\rm e}^{6\,{\frac {r_{{t}}}{\ell}}}}}}}{
\ell}^{-1} \\ \nonumber && -2/3\, \left( {{\rm e}^{-2\,{\frac {r_{{t}}}{\ell}}}}-{\xi}^{2}{
{\rm e}^{4\,{\frac {r_{{t}}}{\ell}}}} \right) \nonumber\\&&\times \left( 2\,{{\rm e}^{-2\,{
\frac {r_{{t}}}{\ell}}}}+{\xi}^{2}{{\rm e}^{4\,{\frac {r_{{t}}}{\ell}}}}+{
{\rm e}^{10\,{\frac {r_{{t}}}{\ell}}}}{\xi}^{4}+2\,{\xi}^{6}{{\rm e}^{16
\,{\frac {r_{{t}}}{\ell}}}} \right) \nonumber \\ && \times {\varphi}^{4} \left( 1+{\xi}^{2}{{\rm e}
^{6\,{\frac {r_{{t}}}{\ell}}}} \right) ^{-8/3}{\ell}^{-3}\nonumber \\ && +\mathcal{O}\left( {\varphi}^{6
} \right)\,.
\end{eqnarray}
We can compute $\ell$ using Eq.~(\ref{l}) and then we can find $r_{t}$  from it.
Finally, we evaluate the volume, and we get the following holographic complexity for entangled cylinder:
\begin{eqnarray}
\Delta\mathcal{C}_A &=&  \frac{-2\,L }{16\pi G_{d+1}}\Big[
\left( \int _{0}^{\phi_{{\infty }}}\! \left( -1+{{\rm e}^{
2/3\,{\frac {{\phi}^{2} \left( 3\,{\ell}^{2}+4\,{\phi}^{2} \right) }{{\ell}^
{4}}}}} \right) {d\phi}  \right) \\ \nonumber && +\mathcal{O} ( {
\xi}^{2})  
\Big].
\end{eqnarray}
It  may be noted that the holographic complexity for the entangled cylinder again depends on the size of the system. 
However, the fidelity susceptibility  does not have such a dependence, as it is calculated for the full system. 
It will be interesting to use this result to understand the behavior of the fidelity susceptibility  for an Abelian Higgs field, as 
the Abelian Higgs field are dual to such solutions. 

\section{Inhomogeneous Backgrounds}\label{inhom}
 Another interesting way to study these holographic quantities is for example, to consider metric with different kind 
 of background as those with inhomogeneous \cite{Bhattacharya:2012mi}. In fact, it has been demonstrated using this solution that
  the entanglement entropy for a very small subsystem obeys a property which is analogous to the first law of thermodynamics 
  when  the system is excited.  It has also been demonstrated that the 
  AdS plane waves describe simple backgrounds which are dual to anisotropically excited systems with energy fluxes \cite{systems}. 
 An inhomogeneous   background has been used to holographically calculate conductivity  \cite{system}. 
 It has been demonstrated that the Drude-like peak and a delta function with a negative weight occur for the real part of this conductivity. 
  Thus, it will be interesting to analyse the fidelity susceptibility and holographic complexity for such a solution, 
  and use it understand the behavior of such systems. 
 
Thus, we will use the    inhomogeneous backgrounds, and  the metric for such  a background can be written as follows \cite{Bhattacharya:2012mi}
\begin{eqnarray}
	&&ds^2=\frac{R^2}{z^2}\Big[-f(z)dt^2+g(r,z)dz^2+dr^2+r^2d\Omega_{d-1}\Big]\,,\label{g1}
\end{eqnarray}
here $g(r,z)=1+m(1+ar+br^2)z^d$ with $m\ll1$, is gravity dual to $CFT_d$.
The entangled region is a round sphere radius $r=\ell$. We parametrize the region by $A=\{t=0,z=z(r)\}$ and hence the area functional becomes
\begin{eqnarray}
	&&Area=\Omega_{d-1} R^d\int \Big(\frac{r^{d-1}}{z^d}
	\Big) \Big[g(r,z)z'^2+1
	\Big]^{1/2}dr\,,
	\label{A1}
\end{eqnarray}
where primes denote derivatives with respect to $r$. The Euler-Lagrange equation for the above area functional is given by
\begin{eqnarray}
	2rzz''&=& \big( -2\,dz \big( r \big) -2\,{z}^{d}m\,a\,d\,r\,z \big( r \big) +2\,
	z \big( r \big) \nonumber \\  &&  -2\,{r}^{2}m{z}^{d}bdz \big( r \big) -2\,d{z}^{d
	}mz \big( r \big) +{z}^{d}marz \big( r \big) \big) {z'}^{3} \nonumber\\ && +\big( 2\,{z}^{d}mz
	\big( r \big)  \big) {z'}^{3}-2\,d{z'}^{2}r+ \big( 2\,z \big( r
	\big) \nonumber\\  &&  -4\,{r}^{2}m{z}^{d}bz \big( r \big) -2\,{z}^{d}m\,a\,r\,z
	\big( r \big) -2\,dz \big( r \big)  \big) z'\nonumber\\ &&-2\,dr+2\,{r}^{2}d
	{z}^{d}ma+2\,d{z}^{d}mr+2\,{r}^{3}d{z}^{d}mb\,.
\end{eqnarray}
Here, in order to obtain a series solution for $z(r)$, we can suppose that $m\ll1$. In addition, we can 
suppose the boundary conditions are $z(0)=z_t$ and $z'(0)=0$, giving us  
\begin{eqnarray}
	z \left( r \right) &=& z_{{t}}+\,{\frac { \left( m{z_{{t}}^{n}-1}
			\right) {r}^{2}}{2z_{{t}}}}-\,{\frac {m{z_{{t}}}^{n}a \left( m
			{z_{{t}}}^{n} -n-1\right) {r}^{3}}{3z_{{t}} \left( n+1 \right) }}\nonumber \\ && 
			+\mathcal{O}(r^4)\,.\label{zr}
\end{eqnarray}
The volume functions reads as follows
\begin{eqnarray}
	&&V=\Omega_{d-1} R^{d+1}\int_{0}^{r_{t}}r^{d-1}dr\int_{0}^{z(r)}\frac{\sqrt{g(r,z)}}{z^{d+1}}dz.
\end{eqnarray}
Using the solution (\ref{zr}) and the approximation $m\ll1$, the finite part of the
volume functional which is obtained by subtracting the pure AdS$_{d+2}$ part from the massive one will be
\begin{eqnarray}
	&&V=-\frac{mz_t \Omega_{d-1} R^{d+1}}{3}\int_{0}^{r_t}\sum c_n r^n(r^2-2 z_t^2)^{k(n)-\frac{1}{2}}dr\,.\label{VVV1}
\end{eqnarray}
The fidelity susceptibility is proportional to the finite part of the following integral,
\begin{eqnarray}
	\mbox{V}_{max}&=&\Omega_{d-1} R^{d+1}\int_{0}^{r_{\infty}}r^{d-1}dr\int_{0}^{z_{\infty}}\frac{\sqrt{g(r,z)}}{z^{d+1}}dz
	\nonumber \\ &\approx& \frac{m b\Omega_{d-1} R^{d+1}}{2(d+2)} r_{\infty}^{d+2}\ln(z_{\infty})\,.\label{VVV2}
\end{eqnarray}
Here $z_{\infty}$ and $r_{\infty}$ are UV cutoff values.  It is important to mention that these two volumes (\ref{VVV1}) and (\ref{VVV2}), 
are  different. Furthermore, the fidelity susceptibility does not scale with the size of the subsystem, as it is calculated for the whole system. 
It can be used to understand the behavior of fidelity susceptibility  a certain CFT$_d$, which is dual to such a solution \cite{Bhattacharya:2012mi}. 
We have also calculated the holographic complexity of this solution, and this was done by using the same surface, which would be used 
to calculate the holographic entanglement entropy of this system.

\section{Hyperscaling Violating Backgrounds }\label{hypers} A hyperscaling geometry   occurs in theories with a 
entropy-temperature relationship given by $S\sim T^{d/z}$, where $d$ is
 the dimension of the space-time and $z$ it is known as a dynamical critical exponent. In other words, these theories have free energy scales determined 
 by there dimensions \cite{hyper1, hyper2, hyper3}. The scaling  behaviors of the mutual information during a process of thermalization
 of such solution has been studied \cite{hype5}. It was demonstrated in this study that   
   during the thermalization process, the dynamical exponent can be used to  obtain   the general time scaling behavior of mutual information. 
 Furthermore, it was demonstrated that the    scaling violating parameter can be employed to define an effective dimension. 
 The DC and Hall conductivity for strange metal has also been studied holographically using such backgrounds \cite{hype6}. 
 This is because such solutions can be used to obtain linear-T resistivity and quadratic-T inverse Hall angle. 
 Now we will analyse   fidelity susceptibility for a hyperscaling violating background, and it will be important 
 to understand the behavior of fidelity susceptibility for strange metals. We will also use a different volume to also calculate the holographic 
 complexity for such systems. 

 Thus, we can start from 
 a non-relativistic hyperscaling violating geometry, and this geometry can be described using the  following metric \cite{hyper1, hyper2, hyper3}
\begin{eqnarray}
	&&ds^2=\frac{R^2}{r^2}\Big[-r^{
		\frac{-2(d-1)(z-1)}
		{d-1-\theta}}dt^2+r^{\frac{2\theta}{d-1-\theta}}
	dr^2+dx_i^2\Big]\,.
	\label{g2}
\end{eqnarray} 
{Here, the index $i=1,2,..,d$ denotes the coordinates for the flat spatial part of the metric and $z$ and $\theta$ are
the dynamical and hyperscaling violating exponent respectively. Moreover, $\theta$ can be interpreted as a the dimension of a zero-energy excitations 
momentum-space surface. Clearly, Lifshitz theories arises when we take $\theta=0$ and $z\neq1$,
whereas CFT are recovered with $\theta=0$ and $z = 1$}. For this space-time, the entangled region can be parametrized to
$A=\{x_1=x_1(r),x_{2,3,..,d}=L\}$, and then the area functional is given by
\begin{eqnarray}
	&&Area=L^{d-1}\int \Big(\frac{R}{r}
	\Big)^d r^{d}\Big[r^{\frac{2\theta}{d-1-\theta}}+(x_1)'^2
	\Big]^{1/2} dr
	\label{A2}
\end{eqnarray}
where primes denotes differentiation with respect to $r$. The functional (\ref{A2}) does not depend on $x_1$, so that the first integral exists,
\begin{eqnarray}
	&&\frac{x_1'}{\Big[r^{\frac{2\theta}{d-1-\theta}}+(x_1)'^2
		\Big]^{1/2} }=\Big(\frac{r}{r_{*}}
	\Big)^d \,,\label{x1r}
\end{eqnarray}
where $x'_{1}(r_{*})=\infty$. The volume can be obtained by
\begin{eqnarray}
	&&V=L^{d-1}R^{d+1}
	\int_{0}^{r_*} r^{\frac{\theta}{d-1-\theta}-(d
		+1)}x_1(r)dr\label{V2}\,,
\end{eqnarray}
where $r_*$ can be found from the total length of the entangled region, which is
\begin{eqnarray}
	&&\ell=2 r_{*}^{\frac{d-1}{d-1-\theta}}\int_{0}^{1}
	\frac{\xi^{d+\frac{\theta}{d-1-\theta}}}{\sqrt{1-\xi^{2d} 
		}}d\xi \,,\ \ \xi=\frac{r}{r_*}\,.
	\end{eqnarray}
From (\ref{x1r}), we can find the following solution
\begin{eqnarray}
	x_1(r)&=&\frac{2r_{*}^{\frac{d-1}{d-1-\theta}}}{ \left( d-1-\theta \right)  \left( d \left( d-1 \right) -\theta 
	\left( d-1 \right) +d-1-\theta \right) } \nonumber\\&& \times \left( \frac {r}{r_*} \right) ^{2 d B }
		{\mbox{$_2$F$_1$}\Big(A, B; C; D \Big)}\,\nonumber\\
		\label{x1sol}
	\end{eqnarray}
	where
	\begin{eqnarray}
	A&=&  \frac{1}{2},\nonumber  \\ 
	B&=& \frac{1}{2}\,{\frac {d \left( d-1 \right) -\theta \left( d-1 \right)
            +d-1-\theta}{ \left( d-1-\theta \right) d}} \nonumber\\  
            C &=& \frac{1}{2}\,{\frac {2\,{d}^{2}-d-2\,d\theta+d 
		\left( d-1 \right) -\theta \left( d-1 \right) -1-\theta}{ \left( d-1-\theta \right) d}}\nonumber  \\ 
		D &=& - 
		\left( \frac {r}{r_*}  \right) ^{2\,d}
	\end{eqnarray}
where $_{2}F_1(A,B;C;D)$ denotes the first hypergeometric function. By replacing (\ref{x1sol}) in (\ref{V2}), we find that the volume is
	\begin{eqnarray}
		&&V=L^{d-1}R^{d+1} r_{*}^{\frac{\theta+d-1}{d-1-\theta}-(d
			+1)}
		\int_{0}^{1} \xi^{\frac{\theta}{d-1-\theta}-(d
			+1)}x_1(\xi)d\xi\,,
	\end{eqnarray}
which can be simplified to the form
	\begin{eqnarray}
		&&V=L^{d-1}R^{d+1}r_{*}^{\frac{\theta+d-1}{d-1-\theta}-(d
			+1)}N(d,\theta)\label{Vhypcomplex}\,.
	\end{eqnarray}
This quantity is a number since $\ell\sim  r_{*}^{\frac{d-1}{d-1-\theta}}$. Thus, holographic complexity becomes
	\begin{eqnarray}
		&&\mathcal{C}=\frac{L^{d-1}R^{d} \ell^{(\theta-(d
				+1)(d-1-\theta))(d-1)}N(d,\theta)}{8\pi G}\,.
	\end{eqnarray}
Finally, the fidelity susceptibility is proportional to the finite part of the following integral
	\begin{eqnarray}
		 \mbox{V}_{max} &=& R^{d+1}L^d\int_{\epsilon}^{r_{\infty}}r^{\frac{\theta}{d-1-\theta}-(d+1)}dr
		\nonumber \\ &\approx& \Big(\frac{R^{d+1}L^d}{\frac{\theta}{d-1-\theta}-d}\Big)r_{\infty}^{\frac{\theta}{d-1-\theta}-d}.  
	\end{eqnarray}
Here, we can see that $ \theta\neq\frac{d(d-1)}{d+1}$ and then the above volume expression is totally different than Eq. (\ref{Vhypcomplex}). 
It may be noted that no trace of Lifshitz exponent $z$ appear in the volumes. Thus, both  the fidelity susceptibility
and holographic complexity will not depend on the Lifshitz exponent $z$.  Furthermore, the holographic complexity also depends 
on the size of the subsystem, even for these backgrounds. However, no such dependence is observed in the fidelity susceptibility, 
as it is calculated for the full system.

\section{Conclusion} \label{conclusion}
The laws of physics can be represented in terms of the ability of an observer to process relevant information. The information theory 
deal with the ability of an observer to process information. It is important to know how much information is lost during such a process, 
and how difficult is it for an observer to process the relevant information during such a process. 
Just as the entropy quantifies the abstract idea of the loss of information, complexity quantifies the idea of the difficulty to 
process that information. It is possible to use the AdS/CFT correspondence to calculate the entanglement entropy of a field theory holographically 
from the bulk geometry dual to such a field theory. This is done by calculating the area in the bulk, as the area in the bulk geometry 
is dual to the holographic entanglement entropy of the boundary theory. Recently, it has been proposed that it is also possible to calculate 
the complexity of a system  holographic, as it is dual to a volume in the bulk. As there are many ways to define a volume in the bulk, many 
different proposals for the complexity have been proposed. If the maximum volume is used, then we obtain the fidelity susceptibility $\Delta \chi_F$. 
However, if the same surface that was used to calculate the entanglement entropy is used, then we obtain a new quantity which is called 
the holographic complexity $\Delta \mathcal{C}$.
In this paper, we
calculate both these quantities for a variety of deformed AdS solutions. 

We calculate it for an AdS black hole,  Janus solution, a solution with  
  cylindrical symmetry, inhomogeneous backgrounds  and hyperscaling violating  backgrounds.
It was observed that most of these geometries are dual to interesting field theories. 
Thus, it was important to calculate and analyze the behavior of holographic complexity 
and fidelity susceptibility for such bulk geometries, as these results can be used 
to understand the behavior of the boundary field theory. 
Furthermore, these geometries where interesting deformations of AdS spacetime, and 
certain universal features were observed to occur in all these different geometries. 
 It was observed that as the holographic complexity depended on the size of the subsystem, 
 and the fidelity susceptibility did not depend on any such size. 
 These observation did not depend on the 
  kind of deformation of the AdS spacetime, and thus seems to be a universal feature 
  of all such deformations. It is also expected to occur as 
  the  holographic complexity was calculated for a subsystem, 
 so it depended on the size of the subsystem. However, as the 
  fidelity susceptibility was calculated for the full system, 
  it did not depend on the size of the subsystem.

It may be noted that these deformed AdS backgrounds are dual to interesting field theories, and many of these field theories have important
condensed matter applications. Thus, we can use the results of this paper, to obtain the 
fidelity susceptibility for those field theories. In fact, many of those theories can be represented by a many-body system. 
 The quantum mechanical 
  Hamiltonian  for such a system,  can be written as  $H(\lambda)=H_0+\lambda H_I$,
where $\lambda$ is an external excitation 
parameter \cite{r6,r7, r8}. It is possible to 
  diagonalize this Hamiltonian by an appropriate set of orthonormal eigenstates 
$\ket{n}$ and eigenvalues $E_m(\lambda)$,  
$ H(\lambda)\ket{n(\lambda)}=E_n(\lambda)\ket{n(\lambda)}.
$  Furthermore, for any two states   $\lambda$ and $\lambda'=\lambda+\delta\lambda$ (which are close to each other), it is also possible to 
 define $F(\lambda,\lambda+\delta\lambda)
=1-\frac{\delta\lambda^2}{2}\chi_F(\lambda)+\mathcal{O}(\delta\lambda^4)
$. Now   the fidelity susceptibility of this system is denoted by   $\chi_F(\lambda)$  \cite{r6,r7, r8}.
It is possible to estimate this  quantity $\chi_F(\lambda)$    holographically as
$
\chi_F(\lambda)=complexity
$ when  $V = V_{max}$ \cite {r5}. 

We have calculated fidelity susceptibility for various bulk solutions, and these bulk solutions 
are dual to interesting boundary theories. Hence, the results of  paper can be used to understand the behavior 
the fidelity susceptibility for those boundary theories, which are dual to the bulk solution analysed in this paper.
We would also like to comment, that at present it is not clear what quantity does holographic complexity represent in the 
boundary theory. It may be a new quantity, which might be closely related to the holographic entanglement entropy, as it is calculated 
using the same surface which is used to calculate the holographic entanglement entropy. It will be interesting to analyse the 
relation between the holographic complexity and holographic entanglement entropy, to understand the implications of the 
holographic complexity for the boundary theory. We have analysed both holographic complexity and fidelity susceptibility
for various solution in this paper, and it would also be interesting to understand the relation between the holographic complexity 
and  fidelity susceptibility, and this might also lead to some understanding of the use of holographic complexity for the boundary theory. 
However, as both these proposals have only been recently made, it was important to apply them to various different deformed AdS solutions, 
and this is what we have done in this paper. 
 
 A  connection has been established  between the holographic entanglement entropy and the quantum phase transition 
 in a lattice-deformed Einstein-Maxwell-Dilaton theory \cite{s2}. In fact, in this study   backgrounds exhibiting metal-insulator transitions
 have been constructed. Furthermore, it has been demonstrated that for these backgrounds  
 both metallic phase and insulating phase have vanishing entropy density,  
 in zero temperature limit. It would be interesting to analyse   holographic complexity and fidelity susceptibility for such backgrounds, 
 and thus use them to study the behavior of metal-insulator transition. 
The holographic phase transition with dark matter sector in the AdS black hole background has also been studied 
\cite{s1}.  It was observed that the  properties of different phases of this system can be 
 obtained from the holographic  entanglement entropy for this system. It would be interesting to analyse 
 the holographic complexity and fidelity susceptibility for such a system. 
\section{Acknowledgments}  SB is supported by the Comisi{\'o}n Nacional de Investigaci{\'o}n Cient{\'{\i}}fica y Tecnol{\'o}gica (Becas Chile Grant No.~72150066). We thank "Seyed Ali Hosseini Mansoori" for correspondence/discussion on results presented in the section (2) of the early draft of this work .


\end{document}